\def\a{\alpha}\def\b{\beta}
 \documentstyle[11pt]{article} 
\begin{document}
\begin{titlepage}
\title{Critique of weakly decohering and linearly positive histories}
\author{Lajos Di\'osi
\thanks{E-mail: diosi@rmki.kfki.hu}\\
KFKI Research Institute for Particle and Nuclear Physics\\
H-1525 Budapest 114, POB 49, Hungary\\\\
{\it bulletin board ref.: gr-qc/9409017}}
\maketitle
\begin{abstract}
The usual composition rule of independent systems, as applied to
decoherent histories or to linearly positive
histories, requires (at least) medium decoherence and, consequently,
a second constraint for the linearly positive histories of
Goldstein and Page. Other plausible classical features, valid for
medium decoherence, seem anyhow to be missed by linear positive histories.
\end{abstract}
\end{titlepage}

{\it Introduction.}
The issue of assigning probabilities to sequences of quantum events
consistently, without referring to the collapses of von Neumann,
has promoted a series of ideas. Among them, the theory of
consistent (in a broad sense) {\it histories} has developed numerous
conditions of {\it decoherence} \cite{Gri84,Omn92,GMH93,DowHal92,Hal94}.
In the first part, we show that {\it weak} decoherence \cite{GMH93}
may contradict to the usual composition rule
of independent systems. At least {\it medium}
decoherence \cite{GMH93} must be imposed to remove the issue.
For the same reason, {\it linearly positive} histories of Goldstein and Page
\cite{GolPag94}
claiming to generalize decoherent histories need an additional
strong constraint.
In the second part, we shall propose a new criterion of {\it dynamical
classicality}. It holds for decoherent histories while it may question
the robust generalization of decoherence proposed by Goldstein and Page.

{\it Histories, probabilities.}
Histories are usually defined by a time-ordered sequence of events:
\begin{equation}
C_\a=P_{\a_n}(t_n)\dots P_{\a_1}(t_1)
\end{equation}
along with a Heisenberg state $\rho$,
where $P_{\a_k}(t_k)$ are exhaustive and exclusive sets of hermitian
projectors assigned to time $t_k$ for each $k=1,\dots,n$ in turn.
In the theory of decoherent histories, one defines the decoherence functional
\begin{equation}
D(\a^\prime,\a)=\langle C_{\a^\prime}^\dagger C_\a \rangle_\rho
\end{equation}
where $<>_\rho$ stands for expectation values in quantum state with
density operator $\rho$.
Weak decoherence imposes the condition
\begin{equation}
Re D(\a^\prime,\a)=0~~~~~for~all~\a\neq\a^\prime,
\end{equation}
and the following probability is assigned to each history of the decohering
family:
\begin{equation}
p_\a=D(\a,\a).
\end{equation}

In the proposal of Goldstein and Page \cite{GolPag94},
the expectation value of the class operator (1) itself is introduced:
\begin{equation}
\pi_\a=\langle C_\a \rangle_\rho,
\end{equation}
and the following condition is imposed on it:
\begin{equation}
Re\pi_\a \geq 0~~~~~for~all~\a.
\end{equation}
Then $C_\a$'s generate a family of linearly positive histories,
each history has the following probability:
\begin{equation}
p_\a=Re\pi_\a.
\end{equation}
Goldstein and Page have proved that weakly  decohering families form
a subset of all linearly positive families, and also
the probability assignment (7) becomes identical with (4) for them. Linear
positivity has been claimed a robust generalization of weak decoherence.

{\it Test of composition rule.} Assume two independent quantum systems
$A$ and $B$ with density operators $\rho^A,\rho^B$, respectively.
Let us assume that the class operators $C_\a^A,C_\b^B$ generate
separate families of weakly decohering histories for $A$ and $B$ respectively.
In ordinary quantum theory a trivial composition of two independent
systems is always possible. In our case the composed system's density
operator is the direct product $\rho^A\otimes\rho^B$.
It is plausible to expect that the class operators
$\{C_\a^A\otimes C_\b^B\}$ will generate a family of
weakly decoherent histories
for the composed system. This latter's decoherence functional factorizes;
in obvious notations this reads:
\begin{equation}
D^{AB}(\a^\prime\b^\prime,\a\b)=D^A(\a^\prime,\a)D^B(\b^\prime,\b).
\end{equation}
It is easy to see that the weak decoherence condition (3) for the
separate systems A and B does not imply the fulfilment of the same
condition for the composed system. To this end, on should replace the
weak decoherence (3) by the stronger medium decoherence condition:
\begin{equation}
D(\a^\prime,\a)=0~~~~~for~all~\a\neq\a^\prime.
\end{equation}
For medium decoherence the expected composition rule of histories
of independent systems becomes valid.

Quite the same reasons warn that the positivity condition
of Goldstein and Page is insufficient. The expectation value (5) of the
composed system's class operator factorizes:
\begin{equation}
\pi_{\a\b}^{AB}=\pi_\a^A\pi_\b^B.
\end{equation}
The positivity (6) of $\pi_\a^A$ and $\pi_\b^B$ does not assure the
positivity of $\pi_{\a\b}^{AB}$.
It is straightforward to replace the positivity condition (6) by the much
stronger one:
\begin{equation}
Re\pi_\a \geq 0,~~~~~Im\pi_a=0~~~~~~for~all~\a.
\end{equation}
In the remaining part, we mean this "medium" version
when speaking about linearly positive histories.

{\it Test of dynamical classicality.}
Assume some external forces leading to the following change of the
Hamiltonian:
\begin{equation}
\delta H(t)=\sum_{k=1}^n\delta(t-t_k)
        \sum_{\a_k}\lambda_{\a_k}(t_k)P_{\a_k}(t_k)
\end{equation}
where the $\lambda$'s are real coupling constants.
This interaction influences any dynamical variable ${\cal O}$
via the Heisenberg equation of motion $d{\cal O}/dt=i[\delta H,{\cal O}]$.
In the general case, the interaction Hamiltonian (12) does not commute with
all projectors a history is made of, c.f. Eq.(1).
If, however, the  history belongs
to a decohering (or, alternatively, linearly positive) family, then all
projectors the given history is made of are thought to have
definite c-number values $0$ or $1$. On a particular subspace,
the history projectors acts as c-numbers and they are
expected to commute with the Hamiltonian (12). So, dynamical classicality
of decohering (linearly positive) histories might require their
invariance against the perturbation by the interaction (12).
We are going to show that this invariance holds for decohering histories and
fails for linearly positive ones.

For simplicity, we assume that only one interaction term on the R.H.S.
of Eq.~(12) has been switched on, say at $t=t_k, 1<k<n$.
Let us introduce the unitary operator corresponding to the interaction:
\begin{equation}
U(t_k)=\exp\left(-i\sum_{\a_k}\lambda_{\a_k}(t_k)P_{\a_k}(t_k)\right).
\end{equation}
According to Heisenberg dynamics, the class operators (1) become different:
\begin{equation}
C_\a=U^\dagger(t_k)P_{\a_n}(t_n)\dots P_{\a_k}(t_k)U(t_k)\dots P_{\a_1}(t_1).
\end{equation}
The decoherence functional (2) takes  modified form:
\begin{equation}
D(\a^\prime,\a)=\left\langle
P_{\a_1^\prime}\dots U^\dagger(t_k)P_{\a_k^\prime}(t_k)
\dots P_{\a_n^\prime}(t_n)P_{\a_n}(t_n)\dots
P_{\a_k}(t_k)U(t_k)\dots P_{\a_1} \right\rangle_\rho
\end{equation}
while the expectation value (5) of the class operator itself changes for
\begin{equation}
\pi_\a=\left\langle
U^\dagger(t_k)P_{\a_n}(t_n)\dots P_{\a_k}(t_k)U(t_k)\dots P_{\a_1}(t_1)
       \right\rangle_\rho.
\end{equation}
{}From the medium decoherence condition (9) [but not from the weak one (3)],
it is easy to prove that the expression (15) reduces to (2),
i.e., decoherent histories are really invariant against the
external perturbations (12). Such invariance is hardly
possible to prove for linearly positive histories: even the
medium constraints (11) may well be violated by the Eq.~(16).

{\it Conclusion.}
We have shown that the complex decoherence functional must be diagonal since
the diagonality of its real part in itself is insufficient to assign
probabilities to quantum histories in a consistent way. Similarly, the
conditions for linearly positive histories, assumed to generalize
weak decoherence, must be much more restrictive than it was thought earlier.
We stated a certain new criterion of dynamical classicality which is admitted
by medium decoherence and missed by linear positivity. Independently of
the ultimate value of the new criterion, one finds medium decoherence
more favorable than its extension to linearly positive histories.
Medium decoherence is equivalent to von Neumann formalism, as emphasized
in Ref. \cite{Dio92}; in fact, this assures the ultimate consistency of
decoherent histories. Nothing but von Neumann probabilities seem to exist.
Any extension of them (weak decoherence, linear positivity)
should be at least doubted and carefully tested.

This work was supported by the grant OTKA No. 1822/1991.
\bigskip



\begin{thebibliography}{99}
\bibitem{Gri84} R.B. Griffiths, J. Stat. Phys. 36 (1984) 219.
\bibitem{Omn92} R. Omn\`es, Rev. Mod. Phys. 64 (1992) 339.
\bibitem{GMH93} M. Gell-Mann and J.B. Hartle, Phys. Rev. D 47 (1993) 3345.
\bibitem{DowHal92}H.F. Dowker and J.J. Halliwell, Phys. Rev. D 46 (1992) 1580.
\bibitem{Hal94} J.J.Halliwell, in {\it Stochastic Evolution of Quantum States
in Open Systems and Measurement Processes}, eds. L. Di\'osi and B. Luk\'acs
(World Scientific, Singapore, 1994), gr-qc/9308005.
\bibitem{GolPag94} S. Goldstein and D.N.Page, preprint Alberta-Thy-43-93; also
as \hbox{gr-qc/9403055}.
\bibitem{Dio92} L. Di\'osi, Phys. Lett. 280B (1992) 71.
\end{thebibliography}
\end{document}